# The Windfall Clause

Distributing the Benefits of AI for the Common Good

Cullen O'Keefe
Centre for the Governance of AI
Future of Humanity Institute
University of Oxford
cullen@openai.com

Peter Cihon
Centre for the Governance of AI
Future of Humanity Institute
University of Oxford

Ben Garfinkel
Centre for the Governance of AI
Future of Humanity Institute
University of Oxford

Carrick Flynn
Centre for the Governance of AI
Future of Humanity Institute
University of Oxford

Jade Leung
Centre for the Governance of AI
Future of Humanity Institute
University of Oxford
jade.leung@governance.ai

Allan Dafoe
Centre for the Governance of AI
Future of Humanity Institute
University of Oxford

**ABSTRACT**

As the transformative potential of AI has become increasingly salient as a matter of public and political interest, there has been growing discussion about the need to ensure that AI broadly benefits humanity. This in turn has spurred debate on the social responsibilities of large technology companies to serve the interests of society at large. In response, ethical principles and codes of conduct have been proposed to meet the escalating demand for this responsibility to be taken seriously. As yet, however, few institutional innovations have been suggested to translate this responsibility into legal commitments which apply to companies positioned to reap large financial gains from the development and use of AI. This paper offers one potentially attractive tool for addressing such issues: the Windfall Clause, which is an *ex ante* commitment by AI firms to donate a significant amount of any eventual extremely large profits. By this we mean an early commitment that profits that a firm could not earn without achieving fundamental, economically transformative breakthroughs in AI capabilities will be donated to benefit humanity broadly, with particular attention towards mitigating any downsides from deployment of windfall-generating AI.

**KEYWORDS:** Automation, Inequality, Future of Work

## 1 Introduction

As the transformative potential of AI has become increasingly salient as a matter of public and political interest, there has been growing discussion about the need to ensure that AI broadly benefits humanity.[1] This in turn has spurred debate on the social responsibilities of large technology companies to serve the interests of society at large. In response, ethical principles and codes of conduct have been proposed to meet the escalating demand for this responsibility to be taken seriously. As yet, however, few institutional innovations have been suggested to translate this responsibility into legal commitments which apply to companies positioned to reap large financial gains from the development and use of AI.

This paper offers one potentially attractive tool for addressing such issues: the Windfall Clause. The Windfall Clause is an *ex ante* commitment by AI firms to donate a significant amount of any eventual extremely large profits. By "extremely large profits," or "windfall," we mean profits that a firm could not earn without achieving fundamental, economically transformative breakthroughs in AI capabilities. By "*ex ante*," we mean that we seek to have the Clause in effect before any individual AI firm has a serious prospect of earning such extremely large profits. "Donate" means, roughly, that the donated portion of the windfall will be used to benefit humanity broadly, with particular attention towards mitigating any downsides from deployment of windfall-generating AI.

We introduce the Windfall Clause not as a bullet-proof solution for the ethical problems associated with windfall-generating AI, but in order to stimulate thought and debate about potential solutions. Much more thought, study, and deliberation will be needed in order to identify the best policy tools.

In this paper we will first outline the motivations for considering a mechanism such as the Windfall Clause. Then, we describe the contents of the Windfall Clause and compare it to analogous corporate philanthropy efforts. We conclude by briefly summarizing both our analysis of the legal permissibility of such a mechanism and reflections on potential concerns with the Windfall Clause.

---

[1] This has centered most visibly on the issue of AI applications beginning to replace human labor in many sectors [2, 10, 13, 15, 25, 32, 35, 39].

## 2  Motivations

The Windfall Clause could advance a number of public policy objectives related to the development of AI. Here, we outline the four most prominent policy motivations.

### 2.1  Addressing Employment Effects

The net effects of AI on aggregate wealth will probably be highly positive due to increased productivity [1]. However, many have argued that AI could lead to a substantial lowering of wages [10], job displacement, and even large-scale elimination of employment opportunities as the structure of the economy changes productivity [8].

One general consideration that supports these concerns is the strong consensus among AI researchers that most, if not all, human work can in theory be automated [1, 26, 46]. Most AI researchers also believe that at some point in the future, progress in AI will make near-complete automation feasible, particularly given that the upper bound of AI capabilities likely exceeds human capabilities [7, 29].[2] Of course, people have unsuccessfully predicted that automation will end human employment many times in the past.[3] We do not mean to imply that such joblessness is inevitable or even probable; only that it is possible given certain assumptions about the future of AI.

If these predictions are realized, then the Windfall Clause could provide an important safety net for those who lose their jobs to automation. For example, the distributed funds could pay for job retraining or subsidize alternative, prosocial activities for which no market yet exists (such as engaging in community service).

### 2.2  Mitigating Income Inequality

Similarly, many worry that automation will increase wealth inequality. Automation increases returns to capital, and capital ownership is highly concentrated [8, 45, 46, 47, 49, 50]. Thus, in the absence of intervention, automation deployed by business could increase the share of income going to shareholders over workers [30, 43, 44].

The Windfall Clause contributes to addressing this by ensuring that extremely large benefits from AI will accrue not just to shareholders and executives of corporate AI labs, but to everyone. In other words, an enforceable Windfall Clause is a credible *ex ante* commitment to prevent extreme income inequality from AI.

### 2.3  Stabilizing Societal Relationships

Advanced AI could radically and rapidly reshape global economics [8]. To the extent that AI disrupts prevailing social and economic conditions, the distribution of windfall profits would afford all beneficiaries a financial buffer. For example, if someone loses her job due to rapid advances in automation,[4] the Windfall Clause could provide a sort of "unemployability insurance" that enables her to weather the transition to a new economy.

This individual stability is desirable in itself, but it could also translate to macro-scale stability. Automatic widespread distribution of gains from advanced AI would decrease the risk of catalyzing popular political pressure for drastic responses to the new economic conditions[5] [32]. Individuals who, thanks to windfall dividends, maintain or even see increases in their standard of living, may be less likely to initiate a destabilizing response [3, 4, 5, 37, 43, 48] and thus threaten the path towards a more prosperous society [6, 19, 34, 36].

### 2.4  Strengthening Prosocial Norms

A number of influential developers, officials, and industry groups have declared their intent to develop AI ethically [9, 17, 18, 27, 33]. Agreeing to be bound by a Windfall Clause provides an immediate and ambitious target for groups interested in promoting broadly beneficial AI and would send a meaningful signal of the signatory's broader support for this goal. This, in turn, would strengthen the emerging international norm of developing AI only for good, making further beneficial uses of AI more likely.

## 3  Contents of the Windfall Clause

### 3.1  Signatory Commitments

Signatories' obligations under the Windfall Clause will ultimately need to be defined according to some mathematical function, even if it is a simple function. In preliminary work we have identified the desiderata shown in table 1 for this "Windfall Function."

Accordingly, we are preliminarily proposing a Windfall Function that would operate similar to an income tax: a higher percentage of marginal revenue would be pledged as profits increased.

Crucially, though, the commitment would scale not with absolute profit levels, but with profit levels relative to the global economy ("elasticity"). Thus, commitments would scale as an individual signatory's profits constituted a greater portion of gross world product (GWP), as shown in table 2.

### 3.2  Distribution Design

After the Windfall Clause has been triggered and contributions have been made in accordance with the Windfall Function, the funds would then need to be distributed. Distributors of Windfall Clause funds would engage in activities that address the public and corporate interests identified above. Ideally, distributors would balance several considerations, including:

1. Effectiveness at promoting the common good and/or addressing the specific downsides of advanced AI;
2. Accountability, so that decision-making is not made on improper bases;
3. Legitimacy, so that people feel the money is being spent fairly and for the common good; and
4. Developers' interests, so that AI developers feel comfortable signing on to the Clause.

---

[2] "Taking the mean over each individual, the aggregate forecast gave a 50% chance of [the arrival of machines that can accomplish, without aid, every task better and more cheaply than humans] occurring within 45 years and a 10% chance of it occurring within 9 years" [29].
[3] See Robots, IGM Forum (February 25th, 2014), perma.cc/N4UD-HE7N?type=image.
[4] This concern already looms large among many workers globally [50].
[5] Most individuals believe that governments are responsible for ensuring workers' success in the face of job automation [50].

Investigating appropriate and desirable options for distributing the Windfall Clause funds is an important area of work which is still in its early stages, and thus could be a fruitful area of future research.

Table 1: Desiderata for "Windfall Function"

| Desideratum | Description | Justification |
| --- | --- | --- |
| Transparency | Determining whether a signatory has met their obligations is easy | Avoids costly disputes and public relations benefits |
| Elasticity | The amount owed under the Clause depends on the size of the global economy when triggered | Maintains focus on relative windfall, not nominal profits |
| Adequacy | The amount owed under the Clause is proportionate to the challenges it seeks to address | Respects the seriousness of the policy motivations for the Clause |
| Nominal Near-Windfall Commitment | Signatories owe small-but-nonzero amounts as they approach windfall profit levels | Demonstrates signatories' intent to be bound; tests effectiveness of Clause |
| Incentive Alignment | Signatories should always have an incentive to earn more profits | Diminishes incentives for evading Clause obligations |
| Competitiveness | Signatories remain competitive with non-signatory developers | Prevents perverse effect of causing non-signatories to out-compete signatories |

Table 2: Commitment scaling based on signatory's profits as portion of gross world product

| Bracket (Profits as Portion of GWP) | Marginal Clause Obligation (Portion of Marginal Profits) |
| --- | --- |
| 0%–0.1% | 0% |
| 0.1%–1% | 1% |
| 1%–10% | 20% |
| 10%–100% | 50% |

## 4 Comparison to Present Philanthropy

As a simplified illustration of how much the Clause might cost a firm in expectation, consider an example wherein a signatory earns $5 trillion (in 2010 dollars) profit from AI in 2060,[6] when the GWP is $268 trillion (again in 2010 dollars). Under the proposed Windfall Function above, such a corporation would be obligated to give $488.12 billion (again in 2010 dollars).

Suppose, optimistically, that a firm in 2019 had a 1% chance of earning such profits by 2060 (and otherwise a 99% chance of earning profits that would not trigger the Windfall Clause). Further suppose, for extreme simplicity, that if the firm achieves windfall profits, under the Clause it will continue to owe $488.12 billion per year forever.[7]

If we discount annually at 10% (the approximate cost of capital for internet software firms) [14, 31], the present cost of such a commitment, if realized, is equal to $64.934 billion (in 2010 dollars). If we further discount that amount by 99% (to account for the probability of not achieving windfall profits and therefore bearing no costs under the Clause), we arrive at an expected present cost of $649.34 million (in 2010 dollars).

To compare this to past corporate philanthropy, 2015's largest donor was Gilead Sciences with $446.7 million [41]. In 2010 dollars, this would be $414.17 million.[8] Note that this historical comparison only includes outright cash donations, not the (opportunity) cost of other forms of CSR [41]. Thus, even under very optimistic assumptions about a signatory's prospects for achieving and maintaining windfall-generating AI, the expected present costs to a signatory are only 60% greater than leading corporate philanthropy efforts today.

---

[6] This year was chosen based on AI researchers' projections of when AI would be able to outperform humans at all economically relevant tasks [29].
[7] In reality, of course, this would depend on how GWP rises relative to the firm's profits. The projections and calculations for such a scenario are far too speculative and beyond the authors' capacity to estimate here. Thus, we use the very simple assumption of perpetual and constant obligations to avoid this speculation. We hope that future work will allow for better estimates of the expected costs of the Windfall Clause.
[8] Calculated using the CPI Inflation Calculator, Bureau of Labor Statistics, perma.cc/4QRE-JAQA (archived Aug. 3, 2019).

## 5 Legal Permissibility

In American corporate law, fiduciary duties require for-profit corporate management to act in the best interests of the corporation. "Corporate expenditures today are judged under the business judgment rule, a standard that accords substantial deference to management's judgment. The fact that a perceived benefit is intangible, noneconomic, or uncertain will not invalidate a corporate expenditure" [24]. Thus, "courts [have] upheld discretionary corporate giving on the theory that donating to charity benefits the corporation," [24][9] as recounted above. Although the Clause is perhaps unique—indeed, somewhat unprecedented—among corporate donations in that it will be very demanding if triggered, for any given firm the expected cost of the commitment should be quite low due to the extremely low probability of obtaining windfall profits. Since the permissibility of corporate actions are evaluated *ex ante* instead of *ex post*[10] [20], this should mean that the Clause is permissible as a matter of corporate law.

A useful analogy can be drawn between the Windfall Clause and stock option compensation, which is incontrovertibly permissible. Like the Windfall Clause, stock option payments have a permissibly low present expected value but can have a much higher value once exercised.[11] For example, in 2005, Facebook issued a large block of stock options to its CEO Mark Zuckerberg [21]. Zuckerberg exercised these options in 2012 [22] and 2013 [23] for a net pre-tax value of over $5.3 billion.[12] If evaluated *ex post* (i.e., after Zuckerberg exercised the options), the 2005 stock options were wildly excessive CEO compensation for a company that earlier that year was evaluated at only $100 million [16]. However, when one evaluates the stock options *ex ante*—as is proper—then they were appropriate. Analogously, the Windfall Clause might appear excessive as a corporate donation if evaluated *ex post*, but it is clearly reasonable if evaluated *ex ante*.

## 6 Two Major Objections

We have spent a lot of time considering potential concerns and flaws of the Windfall Clause. In this section, we reflect on two common concerns.

6.1 "Windfall profits should just be taxed."

Some might object to—or at least be uneasy with—extra-governmental disposition of windfall profits, perhaps arguing instead that the windfall should simply be taxed (if earned by a private individual or organization) and spent by democratically accountable governments [11, 12, 28, 38, 40, 42, 45].

Taxation might be overall preferable to a private Windfall Clause. The Windfall Clause leaves open the possibility of taxation, and could be drafted so as to relieve firms of their responsibilities if combined Clause and tax burdens exceeded some reasonable threshold.

An advantage of the Windfall Clause, however, is that it is more presently actionable than dramatic tax reform since it depends only on convincing individual firms, not political majorities. Further, we hope that promotion of the Windfall Clause or related solutions to the economic problems and opportunities posed by automation will serve as a catalyst for broader exploration of the solution space, including taxation.

However, we also note that the economic effects of automation are likely to transcend political boundaries. Thus, a primary advantage of the Clause over most current forms of taxation is that the Clause could distribute benefits internationally. This could diminish adversarial international competition for the benefits of AI and also lead to more effective uses of the funds given global inequality.[13]

6.2 "Firms will find a way to circumvent their commitments under the Clause."

We do worry that the Clause could be circumvented either by either legal (e.g., contractual loopholes or strategic accounting practices) or extralegal (e.g., intentionally evading Clause obligations) means.

Some of this problem is mitigable—if not completely solvable—by careful contractual and mechanism design. Crafting a Clause with meaningful monitoring and enforcement mechanisms is a significant task and would be important to the success of the Clause.

Most remaining problems would arise from a firm intentionally flouting its legal commitments. Such problems warrant significant attention from the AI policy community as well, but the Clause probably cannot solve them alone.

## 7 Conclusion

Artificial Intelligence may be poised to fundamentally change the nature of the global economy. One possible future is a scenario in which advanced AI services result in unprecedented "windfall" profits that accrue to a very small number of actors. This scenario is highly undesirable, not only because humanity at large will have borne the risk of innovations along the way, but also because the creation of such wealth might be accompanied by mass unemployment and other occurrences that increase human suffering.

---

[9] Citing Kahn v. Sullivan, 594 A.2d 48, 61 (Del. 1991); Theodora Holding Corp. v. Henderson, 257 A.2d 398, 405 (Del. Ch. 1969); A.P. Smith Mfg. Co. v. Barlow, 98 A.2d 581, 590 (N.J. 1953).
[10] See, e.g., In re Tyson Foods, Inc., 919 A.2d 563, 586 (Del. Ch. 2007).
[11] Indeed, since there is no upper bound on stock price, stock options could have arbitrarily high value *ex post*.
[12] All options had an exercise price of $0.06, and Zuckerberg acquired 60 million shares in each transaction [22, 23]. Facebook's stock was valued at $34.03/share for the 2012 transaction. See FB Historical Prices, Yahoo! Finance, perma.cc/GE8A-CBKZ (archived Feb. 7, 2019). Thus, in the 2012 transaction, Zuckerberg paid $3.6 million for $2.0418 billion in stock for a net pre-tax gain of $2.0382 billion. Facebook's stock was valued at $55.12/share for the 2013 transaction. See FB Historical Prices, Yahoo! Finance, perma.cc/JJJ5-VM46 (archived Feb. 7, 2019). Thus, in the 2013 transaction, Zuckerberg paid $3.6 million for $3.3072 billion in stock for a net pre-tax gain of $3.3036 billion.

[13] See "Your Dollar Goes Further Overseas," GiveWell, perma.cc/GNP3-4RHJ (archived Apr. 15, 2019).

The Windfall Clause aims to address the downsides of such a scenario, if the development of truly windfall-generating AI were to occur. With further refinement, the Windfall Clause could be a significant, credible, and tractable way for AI firms to direct their inventions towards the enrichment of humanity generally, and still reap substantial rewards for doing so.

We hope to contribute an ambitious and novel policy proposal to an already rich discussion on this subject. More important than this policy itself, though, we look forward to continuously contributing to broader conversations on the economic promises and challenges of AI, and how to ensure AI benefits humanity as a whole.

## REFERENCES

[1] Philippe Aghion, Benjamin F. Jones, and Charles I. Jones. 2017. *Artificial Intelligence and Economic Growth.* Working Paper No. 23928. National Bureau of Economic Research.

[2] Jonathan P. Allen. 2017. *Technology and Inequality.* (1st. ed.). Springer, New York.

[3] Therese F. Azeng and Thierry U. Yogo. 2013. *Youth Unemployment and Political Instability in Selected Developing Countries.* African Development Bank, Tunis, Tunisia.

[4] Aniruddha Bagchi and Jomon A. Paul. 2018. Youth Unemployment and Terrorism in the MENAP (Middle East, North Africa, Afghanistan, and Pakistan) Region. *Socioecon. Plann. Sci.* 64 (Dec. 2018), 9-20. DOI:https://doi.org/10.1016/j.seps.2017.12.003

[5] Efraim Benmelech, Claude Berrebi, and Esteban F. Klor. 2010. *Economic Conditions and the Quality of Suicide Terrorism.* Working Paper No. 16320. National Bureau of Economic Research.

[6] Bruno Biais and Enrico Perotti. 2002. Machiavellian Privatization. *Amer. Econ. Rev.* 92, 1 (Mar. 2002), 240-258. DOI:https://doi.org/10.1257/000282802760015694

[7] Nick Bostrom. 2014. *Superintelligence: Paths, Dangers, Strategies.* (1st. ed.). Oxford University Press, Oxford.

[8] Nick Bostrom, Allan Dafoe, and Carrick Flynn. Forthcoming. Public Policy and Superintelligent AI: A Vector Field Approach. In *Ethics of Artificial Intelligence.* S. Matthew Liao (Ed.). Oxford University Press, New York.

[9] Sergey Brin. 2017. *2017 Founders' Letter.* Alphabet Investor Relations.

[10] Erik Brynjolfsson and Andrew McAfee. 2014. *The Second Machine Age: Work, Progress, and Prosperity in a Time of Brilliant Technologies.* (1st. ed.). W. W. Norton & Company, New York.

[11] Ed Burmila. 2018. Jeff Bezos, Amazon and Why 'Charity' Is the Wrong Solution. *Rolling Stone* (Feb. 14, 2018).

[12] David Callahan. 2017. *The Givers: Wealth, Power, and Philanthropy in a New Gilded Age.* (1st. ed.). Alfred A. Knopf, New York.

[13] Allan Dafoe. 2018. *AI Governance: A Research Agenda.* Governance of AI Program, Future of Humanity Institute, University of Oxford, Oxford, UK.

[14] Aswath Damodaran. 2019. *Cost of Capital by Sector (Us).* NYU Stern School of Business, New York.

[15] John Danaher. 2019. *Automation and Utopia : Human Flourishing in a World without Work.* (1st. ed.). Harvard University Press, Cambridge, MA.

[16] Dealbook. 2012. Tracking Facebook's Valuation. *N.Y. Times* (Feb. 1, 2012).

[17] DeepMind. 2018. Deepmind Ethics & Society Principles.

[18] Jeffrey Ding. 2018. *Deciphering China's AI Dream.* Future of Humanity Institute.

[19] Denise DiPasquale and Edward L Glaeser. 1999. Incentives and Social Capital: Are Homeowners Better Citizens? *J. Urban Econ.* 45, 2 (Mar. 1999), 354-384. DOI:https://doi.org/10.1006/juec.1998.2098

[20] Frank H. Easterbrook and Daniel R. Fischel. 1996. *The Economic Structure of Corporate Law.* (1st. ed.). Harvard University Press, Cambridge, MA.

[21] Facebook Inc. 2005. 2005 Officers' Stock Plan.

[22] Facebook Inc. 2012. Statement of Changes in Beneficial Ownership (Form 4).

[23] Facebook Inc. 2013. Statement of Changes in Beneficial Ownership (Form 4).

[24] Jill E. Fisch. 1997. Questioning Philanthropy from a Corporate Governance Perspective. *NY Law Sch. Law Rev.* 41 (1997).

[25] Martin R. Ford. 2009. *The Lights in the Tunnel: Automation, Accelerating Technology and the Economy of the Future.* (1st. ed.). Acculant Publishing.

[26] Carl Benedikt Frey and Michael A. Osborne. 2017. The Future of Employment: How Susceptible Are Jobs to Computerisation? *Technol. Forecast. Soc. Change* 114 (Jan. 2017), 254-280. DOI:https://doi.org/10.1016/j.techfore.2016.08.019

[27] Future of Life Institute. 2017. Asilomar AI Principles.

[28] Anand Giridharadas. 2018. *Winners Take All: The Elite Charade of Changing the World.* (1st. ed.). Alfred A. Knopf, New York.

[29] Katja Grace, John Salvatier, Allan Dafoe, Baobao Zhang, and Owain Evans. 2018. When Will AI Exceed Human Performance? Evidence from AI Experts. *J. Artif. Intell. Res.* 62 (Jul. 2018), 729-754. DOI:https://doi.org/10.1613/jair.1.11222

[30] Karen Harris, Austin Kimson, and Andrew Schwedel. 2018. *Labor 2030: The Collision of Demographics, Automation and Inequality.* Bain & Company.

[31] James R. Hitchner. 2010. *Financial Valuation: Applications and Models.* (3rd. ed.). John Wiley & Sons, Hoboken, NJ.

[32] Michael C. Horowitz, Gregory C. Allen, Edoardo Saravalle, Anthony Cho, Kara Frederick, and Paul Scharre. 2018. *Artificial Intelligence and International Security.* CNAS.

[33] IEEE. 2017. Ethically Aligned Design, Version 2.

[34] Saumitra Jha. 2015. Financial Asset Holdings and Political Attitudes: Evidence from Revolutionary England. *Q. J. Econ* 130, 3 (Aug. 2015), 1485-1545. DOI:https://doi.org/10.1093/qje/qjv019

[35] Jerry Kaplan. 2015. *Humans Need Not Apply: A Guide to Wealth and Work in the Age of Artificial Intelligence.* (1st. ed.). Yale University Press, New Haven, CT.

[36] Markku Kaustia, Samuli Knüpfer, and Sami Torstila. 2016. Stock Ownership and Political Behavior: Evidence from Demutualizations. *Manage. Sci.* 62, 4 (Apr. 2016), 945-963. DOI:https://doi.org/10.1287/mnsc.2014.2135

[37] Seymour Martin Lipset. 1959. Some Social Requisites of Democracy: Economic Development and Political Legitimacy. *Amer. Polit. Sci. Rev.* 53, 1 (Mar. 1959), 69-105. DOI:https://doi.org/10.2307/1951731

[38] Annie Lowrey. 2018. Jeff Bezos's $150 Billion Fortune Is a Policy Failure. *Atlantic* (Aug. 1, 2018).

[39] James Manyika, Susan Lund, Michael Chui, Jacques Bughin, Jonathan Woetzel, Parul Batra, Ryan Ko, and Saurabh Sanghvi. 2017. *Jobs Lost, Jobs Gained: Workforce Transitions in a Time of Automation.* McKinsey Global Institute.

[40] Ryan Pevnic. 2016. Philanthropy and Democratic Ideals. In *Philanthropy in Democratic Societies: History, Institutions, Values.* Rob Reich, et al. (Eds.). University of Chicago Press, Chicago, 226-243.

[41] Caroline Preston. 2016. The 20 Most Generous Companies of the Fortune 500. *Fortune* (Jun. 22, 2016).

[42] Rob Reich. 2013. What Are Foundations For?. *BRev* 38 (Mar. 2013), 10-15

[43] Jeremy Rifkin. 1995. *The End of Work: The Decline of the Global Labor Force and the Dawn of the Post-Market Era.* (1st. ed.). G.P. Putnam's Sons, New York.

[44] Jeffrey D. Sachs. 2019. R&D, Structural Transformation, and the Distribution of Income. In *The Economics of Artificial Intelligence: An Agenda.* Ajay Agrawal, et al. (Eds.). University of Chicago Press, Chicago, 329-348.

[45] Jeffrey D. Sachs and Laurence J. Kotlikoff. 2012. *Smart Machines and Long-Term Misery.* Working Paper No. 18629. National Bureau of Economic Research.

[46] Aaron Smith and Janna Anderson. 2014. *AI, Robotics, and the Future of Jobs.* Pew Research Center.

[47] Betsey Stevenson. 2019. Artificial Intelligence, Income, Employment, and Meaning. In *The Economics of Artificial Intelligence: An Agenda.* Ajay Agrawal, et al. (Eds.). University of Chicago Press, Chicago, 189=195.

[48] Daniel Treisman. 2015. Income, Democracy, and Leader Turnover. *Am. J. Polit. Sci.* 59, 4 (Oct. 2015), 927-942. DOI:https://doi.org/10.1111/ajps.12135

[49] Darrell M. West. 2018. *The Future of Work: Robots, AI, and Automation*. (1st. ed.). Brookings Institution Press, Washington, D.C.

[50] Richard Wike and Bruce Stokes. 2018. *In Advanced and Emerging Economies Alike, Worries About Job Automation.* Pew Research Center.